\documentclass[manuscript]{acmart}
\usepackage{xcolor}
\AtBeginDocument{\color{black}}
\usepackage{xcolor}
\AtBeginDocument{\color{black}}
\newcommand{\revise}[1]{\textcolor{black}{#1}}
\usepackage{enumitem}

\begin{document}

\title{Tracing Generative AI in Digital Art: A Longitudinal Study of Chinese Painters’ Attitudes, Practices, and Identity Negotiation}

\author{Yibo Meng}
\authornote{Both authors contributed equally to this research.}
\affiliation{%
  \institution{Health Informatics, Cornell University}
  \city{Ithaca}
  \state{NY}
  \country{USA}
}
\email{mengyb22@tsinghua.org.cn}

\author{Ruiqi Chen}
\authornotemark[1]
\affiliation{%
  \institution{Human Centered Design and Engineering, University of Washington}
  \city{Seattle}
  \state{WA}
  \country{USA}
}
\email{ruiqich@uw.edu}

\author{Zhuoran Lu}
\affiliation{%
  \institution{Human-Computer Interaction, Purdue University}
  \city{West Lafayette}
  \state{IN}
  \country{USA}
}
\email{lu800@purdue.edu}

\author{Shuai Ma}
\affiliation{%
  \institution{Computational Behavior Lab,Aalto University}
  \city{Espoo}
  \country{Finland}
}
\email{mashuai@iscas.ac.cn}

\author{Chengxi Zang}
\affiliation{%
  \institution{Health Informatics, Cornell University}
  \city{Ithaca}
  \state{NY}
  \country{USA}
}
\email{chz4001@med.cornell.edu}

\renewcommand{\shortauthors}{Meng et al.}

\begin{abstract}
  This study presents a five-year longitudinal mixed-methods study of 17 Chinese digital painters, examining how their attitudes and practices evolved in response to generative AI. Our findings reveal a trajectory from resistance and defensiveness, to pragmatic adoption, and ultimately to reflective reconstruction, shaped by strong peer pressures and shifting emotional experiences. Persistent concerns around copyright and creative labor highlight the ongoing negotiation of identity and values. This work contributes by offering rare longitudinal empirical data, advancing a theoretical lens of “identity and value negotiation,” and providing design implications for future human–AI collaborative systems.

\end{abstract}




\maketitle

\section{INTRODUCTION}
The rise of generative artificial intelligence (AI) is rapidly reshaping the landscape of creative practice \cite{shelby2024generative, canet2022dream, chung2022artistic}. Across domains such as illustration, concept design, music, and literature, AI systems are increasingly positioned not just as tools, but as collaborators, consultants, and sometimes even competitors to human creators \cite{cetinic2022understanding, cetinic2022understanding, page2025creative, xu2024application, caramiaux2022explorers}. These systems can produce high-quality outputs at unprecedented speed—offering new sources of inspiration, expanding expressive possibilities, and enhancing creative efficiency. Yet alongside these promises come profound tensions around authorship, originality, and the future of creative labor \cite{kyi2025governance, tang2024exploring, knight2024impact}. Understanding how artists adopt, adapt to, or resist these emerging tools has thus become a pressing concern in Human-Computer Interaction (HCI). As generative systems continue to advance, researchers and designers are faced with an urgent question: how can we build human–AI collaborations that amplify productivity while safeguarding human creativity, agency, and professional identity?

Existing research has begun to explore this space along three main lines: artists’ attitudes toward AI \cite{sikorski2025attitudes, johnston2024understanding, inie2023designing, park2024we}, modes of human–AI collaboration \cite{rezwana2025human, he2023exploring, schecter2025role, xie2025embodied}, and risks related to authorship, credit, and creative labor \cite{kyi2025governance, knight2024impact}. These studies highlight a duality in artists’ views: AI is celebrated as an “engine of inspiration,” capable of fostering divergent ideas and improving efficiency, yet also feared as a threat that may erode professional identity and diminish creative value. However, most prior work relies on cross-sectional surveys or short-term experiments \cite{lu2025designing, he2023exploring, ma2024drawing, lawton2023tool}, offering only static snapshots of a rapidly evolving phenomenon. Such approaches cannot capture how attitudes unfold over time, how collaboration practices shift, or how ethical concerns and value frameworks are reshaped as AI becomes more deeply embedded in creative ecosystems.

\revise{We focus on \textit{digital painters}—artists who primarily work with digital media such as illustration, concept design, and game or animation art—because they represent one of the creative communities most directly affected by generative AI.}
To address this gap, we conducted a five-year longitudinal mixed-methods study (2021–2025) that followed 17 Chinese digital painters through annual surveys and in-depth interviews. By combining quantitative trend analysis with rich qualitative narratives, we captured both shifts in attitudes and the lived stories of how creators negotiated practice and identity. This approach moves beyond the limits of short-term studies, allowing us to examine how relationships with generative AI are continually reconstructed across different stages, social contexts, and professional positions.

Our findings reveal a dynamic trajectory in artists’ responses to generative AI. Quantitative results show a progression from initial resistance and defensiveness, to pragmatic acceptance, and eventually to reflective reconsideration and identity reconstruction. Qualitative analysis further uncovers the lateral forces shaping this arc: peer influence often drove “reluctant adoption,” while emotions shifted from curiosity to anxiety, from excitement to fatigue, and finally toward more measured reflection. Throughout this trajectory, ethical concerns—particularly around copyright, authorship, and the precarity of creative labor—remained consistently salient and grew more pronounced as AI tools became widespread.

This study makes three contributions to HCI. (1) It provides a rare five-year longitudinal dataset that traces the evolving relationship between creators and generative AI, complementing prior cross-sectional work with a temporal perspective. (2) It advances a theoretical lens that frames technology adoption as an ongoing negotiation of identity and values rather than a simple binary of acceptance or rejection. (3) It offers design implications for future creative AI systems: emphasizing user control and transparency, supporting diverse modes of collaboration, and preserving the creative potential of failure and serendipity. Together, these contributions deepen our understanding of how generative AI reshapes creative practice and point toward systems that respect and sustain human creativity.

\section{RELATED WORK}
\subsection{Attitudes toward Generative AI in Creative Practice}
Recent HCI and creativity research has paid increasing attention to how artists perceive, adopt, and contest generative AI tools \cite{bird2024artists,inie2023designing,hertzmann2020computers,koch2020art,tang2024exploring,lovato2024foregrounding}. Many studies emphasize the opportunities that AI brings to creative workflows—highlighting how these systems can accelerate prototyping, expand the space of possible ideas, and facilitate experimentation with new visual styles \cite{xu2024application,he2023exploring,chung2022artistic,lu2025designing,chung2021intersection}. When framed as an “engine of inspiration” or as a collaborative partner rather than a replacement, generative AI is often evaluated more favorably, with artists describing its ability to foster creativity, playfulness, and serendipitous discovery.

However, a parallel body of work documents artists’ ambivalence and resistance \cite{cetinic2022understanding,ch2019art,yang2020re,park2024we}. For example, Kawakami and Venkatagiri \cite{kawakami2024impact} analyzed social media discussions and uncovered widespread anxieties about the devaluation of artistic labor and the erosion of professional opportunities. Similarly, Sikorski et al. \cite{sikorski2025attitudes} surveyed professionals and students in game development, finding particularly negative sentiments among students and specialized artists, driven by fears of displacement and labor precarity. These concerns suggest that even as creators recognize the benefits of generative AI, they often experience it as a disruptive force that challenges established professional identities and norms. Such ambivalence also extends to questions of artistic authorship and public legitimacy. Bird \cite{bird2024artists}, for instance, highlights how artists often navigate competing pressures—to harness AI’s generative power while also resisting narratives that frame it as a creative equal. These tensions emerge not only through direct engagement with tools, but also through broader dynamics of peer reception, community standards, and cultural discourse.

Despite these insights, much of the existing literature captures artists’ attitudes through single-time-point surveys or interviews, offering limited visibility into how views evolve over time. Our work addresses this gap by contributing five years of longitudinal data, tracing how digital painters’ attitudes progressed from early skepticism, to pragmatic adoption, and eventually to critical reconsideration. In doing so, we move beyond static accounts of acceptance or resistance, foregrounding the temporal and sociocultural dynamics that shape long-term engagement with generative AI.

\subsection{Human–AI Collaboration in Creative Work}

Another stream of research has examined how artists’ experiences with generative AI vary depending on the role ascribed to the system—whether framed as an advisor, collaborator, or driver \cite{canet2022dream,johnston2024understanding,shelby2024generative,darabipourshiraz2025ai,cetinic2022understanding}. When positioned as an \textit{advisor}, AI is typically viewed as supporting rather than supplanting human decision-making \cite{stork2024computer,bennett2024painting,twomey2022three,porquet2025copying,moruzzi2024user,mikalonyte2022can}. In this role, AI may offer stylistic references, compositional alternatives, or technical enhancements that improve creative outcomes while preserving human agency and authorship \cite{rezwana2025human,davis2025co,deshpande2024perceptions,ma2024drawing}. For instance, Hu et al. \cite{hu2025designing} identify recurring patterns across HCI studies in which AI functions as a “consultant,” enriching the creative process without undermining ownership. Similarly, Schecter and Richardson \cite{schecter2025role} argue that the advisory framing fosters acceptance by explicitly centering human control.

When cast as a \textit{collaborator}, AI is often described as a co-creative partner that stimulates exploration and creative divergence \cite{lawton2023tool,xie2025embodied,caramiaux2022explorers}. Chung \cite{chung2022artistic}, for example, highlights how systems that capture user intent and allow for iterative refinement are perceived as discovery-oriented collaborators rather than productivity aids. Page et al. \cite{page2025creative} similarly demonstrate that through embodied interaction, artists increasingly engaged with AI as an “inspiration engine”—embracing its capacity to prompt new directions. Crucially, many artists also appropriate AI’s “errors” and unexpected outputs as creative stimuli, treating these moments of unpredictability not as flaws but as openings for serendipitous invention.

By contrast, when AI takes on the role of \textit{driver}, collaboration becomes more fraught. While artists often acknowledge AI’s technical strengths—such as replicating local aesthetic features like color, texture, or shading—they remain skeptical of its ability to convey holistic style, intent, or authorship \cite{porquet2025copying,he2023exploring,johnston2024understanding,mikalonyte2022can,chen2024exploring}. Didion et al. \cite{didion2024did} show that the perceived consistency and predictability of AI outputs shape creators’ sense of authorship: when outputs feel misaligned or overly uniform, artists are less likely to claim them as their own. Schecter and Richardson \cite{schecter2025role} further argue that assigning AI a driver role tends to obscure human contributions and diminish the perceived value of collaboration.

Overall, this body of work illustrates the dual character of human–AI collaboration: AI can extend, amplify, or constrain creative agency depending on how its role is framed. Yet, most existing studies are grounded in short-term evaluations or lab-based prototypes, offering limited insight into how collaborative relationships with AI evolve over time. Our five-year longitudinal study addresses this gap by tracing how digital painters repeatedly renegotiated their relationship with AI—oscillating between curiosity, reliance, resistance, and redefinition as both technological capabilities and social expectations shifted.

\subsection{Problematic Aspects of Generative AI Use}
Beyond explorations of creative potential and collaborative dynamics, a growing body of research has highlighted the risks and structural challenges accompanying generative AI in creative practice \cite{bran2023emerging,brand2021design,cetinic2022understanding}. A central concern relates to copyright and data provenance \cite{daniele2019ai+,amato2019ai}. Prior studies document artists’ fears that their work has been appropriated for training without consent—raising anxieties around creative theft and further eroding the perceived originality of AI-generated outputs \cite{zeilinger2021tactical,kudless2023hierarchies}. Kawakami and Venkatagiri \cite{kawakami2024impact}, for instance, show that these concerns extend beyond independent creators to professional domains such as game development and illustration, where the stakes of authorship and ownership are especially acute.

Building on these issues, concerns around attribution and compensation have shifted focus from training datasets to the downstream distribution and reception of AI-generated content \cite{lima2025public,kawakami2024impact,lovato2024foregrounding,kirova2023ethics}. Kyi et al. \cite{kyi2025governance} synthesize interviews with 20 creative professionals into the “3C” framework—\textit{consent, credit, and compensation}—to capture artists’ multifaceted expectations. While some seek explicit attribution or financial returns, others worry that being credited may associate them with low-quality AI outputs, risking reputational damage. These tensions reveal how recognition in AI-assisted art is not merely a matter of fairness, but a high-stakes negotiation over professional reputation, authorship, and market legitimacy.

Aesthetic concerns further complicate artists’ perceptions of generative AI. Page et al. \cite{page2025creative} report that many creators describe AI-generated images as “overly polished” yet emotionally hollow—employing metaphors of “soullessness” to express their unease. Porquet \cite{porquet2025copying} similarly argues that while AI systems may reproduce superficial stylistic traits such as color palettes or compositional motifs, they fall short in conveying the intentionality and coherence that artists associate with authentic expression. These critiques reflect deeper fears that artistic style itself is being commodified into an extractable resource—divorced from the embodied practices, labor histories, and cultural contexts that grant it meaning \cite{messer2024co,loi2024inflammable}.

Finally, researchers have drawn attention to the precarious labor conditions that AI threatens to exacerbate \cite{ming2024labor,klinova2021ai,zhang2022algorithmic,knight2024impact,munoz2023identity,mako2022emerging}. Lu \cite{lu2025research} demonstrates how the efficiency gains enabled by generative AI have pressured clients to lower compensation rates, disproportionately affecting freelancers and early-career artists. Bird \cite{bird2024artists} likewise finds that many creators adopt AI tools not out of genuine enthusiasm but as a survival strategy in an increasingly competitive market. As Bird notes, artists are often caught in a double bind—compelled to embrace AI to remain relevant, while simultaneously resisting its encroachment to preserve the integrity of their profession.

While these studies underscore urgent concerns surrounding ownership, attribution, aesthetics, and labor, most capture attitudes at a single point in time. What remains underexplored is how such anxieties evolve, intensify, or subside as artists continue to engage with generative AI in everyday practice. Our five-year longitudinal study addresses this gap by tracing how digital painters’ initial concerns transformed into more reflective negotiations over authorship, identity, and value.

\section{METHODOLOGY}
We adopted a five-year longitudinal mixed-methods design (2021–2025) to examine how digital painters’ perceptions of generative AI evolved over time. Longitudinal research is particularly suited for exploring change across time \cite{audulv2023time, saldana2003longitudinal}. At each annual point, we combined quantitative surveys, which offered comparable measures of cognition, usage, and attitudes, with semi-structured interviews, which provided narratives of practice and reflection. Inspired by qualitative longitudinal research traditions and aligned with perspectives that emphasize understanding practice and context rather than isolated outcomes \cite{chalfen1987snapshot, dourish2006implications}, our design enabled us to trace both attitudinal trajectories and the ways painters integrated—or resisted—AI in their workflows.

\subsection{Participants}

We recruited participants in 2021 by posting calls on major Chinese social media platforms, including WeChat, Xiaohongshu, Bilibili, Baidu Tieba, and Weibo. Eligibility criteria required participants to be 18 years or older, have at least one year of digital painting experience (either professional or as a hobby), possess a basic awareness of generative AI (e.g., having heard of or having a general understanding of the concept), and commit to participating in a five-year longitudinal study. Individuals whose primary creative practice was non-digital (e.g., traditional painting, sculpture) or non-illustrative digital work (e.g., 3D modeling, video editing, music production) were excluded to maintain the study’s focus. Those with less than one year of digital painting experience or unwilling to commit to long-term participation were also excluded.

In this study, digital painters refer to artists who primarily create two-dimensional visual works using digital tools such as \textit{Photoshop}, \textit{Clip Studio Paint}, or \textit{Procreate}. Their practice typically spans illustration, concept design, and game or animation art, where they balance artistic expression with client requirements and fast-paced production cycles. This hybrid form of creative labor positions digital painters between independent artistry and commercial design, making them among the creative communities most directly affected by generative AI.

A total of 17 participants were enrolled in 2021, which serves as the baseline demographic profile for this longitudinal study. The sample included 7 men and 10 women, aged 19--34 ($M=26.3$, $SD=4.37$). All participants identified as East Asian. Participants generally had high educational backgrounds (6 postgraduate, 10 undergraduate, 1 high school). Eleven ($N=11$) participants identified as professional digital painters (employed in game studios, publishing houses, or design firms), and six ($N=6$) as hobbyists. Their creative domains included game art, commercial illustration, concept design, anime and graphic novels, and graphic design.

Attrition remained modest across the five-year study. All 17 participants completed the first wave in 2021. Fifteen participants participated in 2022, fourteen in 2023, and thirteen participants completed both the 2024 and 2025 waves. Table 1 summarizes the baseline demographics and annual participation, while Table 3 provides detailed information about participants’ creative domains, client types, and professional status across the five study waves.

This study was reviewed and approved by the Institutional Review Board of the University of Shanghai for Science and Technology (the institution where the first author was affiliated at the commencement of this study) under approval number 20210407. Prior to participating in the study, all participants signed an informed consent form and were fully apprised of the study's objectives, the nature of the data to be collected, and how such data would be utilized. Participants were explicitly informed of their right to withdraw from the study at any time without penalty, as well as their right to request the deletion of their associated data. Given that this study is a five-year longitudinal project, we re-affirmed participants' informed consent prior to each round of data collection to ensure their continued voluntary participation. All interviews were conducted online and, with the participants' consent, were audio-recorded for subsequent transcription and analysis. To safeguard participant privacy and confidentiality, all data underwent anonymization during the transcription and analysis processes, with any personally identifiable information removed. The research data are stored on a secure institutional cloud server, accessible only to authorized members of the research team. Each participant received an annual stipend of RMB 40 (totaling RMB 200 over the five-year period) as compensation for completing the annual questionnaires and interviews.

\begin{table}[htbp]
\centering
\caption{Summary of Participant Demographics (Baseline in 2021, $N=17$).}
\label{tab:participants}
\begin{tabular}{ccccccc}
\toprule
ID & Age (2021) & Gender & Education & Urban/Rural & Years of Drawing & Professional Status \\
\midrule
1 & 22 & F & Bachelor & Urban & 2  & Professional \\
2 & 24 & F & Bachelor & Urban & 6  & Professional \\
3 & 22 & M & Bachelor & Rural & 7  & Professional \\
4 & 19 & M & Master   & Urban & 3  & Non-professional \\
5 & 33 & F & Master   & Urban & 11 & Professional \\
6 & 26 & F & Master   & Urban & 8  & Professional \\
7 & 34 & M & Bachelor & Rural & 8  & Professional (later attrition) \\
8 & 31 & M & Master   & Urban & 7  & Professional (attrition after 2022) \\
9 & 29 & F & Bachelor & Rural & 8  & Professional \\
10 & 26 & M & Bachelor & Urban & 8  & Non-professional $\rightarrow$ Professional \\
11 & 26 & F & High School & Rural & 7 & Professional \\
12 & 25 & F & Master   & Urban & 7  & Non-professional \\
13 & 33 & M & Bachelor & Urban & 8  & Non-professional \\
14 & 22 & F & Bachelor & Rural & 7  & Non-professional \\
15 & 27 & M & Bachelor & Rural & 9  & Professional \\
16 & 24 & F & Bachelor & Rural & 11 & Professional (attrition) \\
17 & 24 & F & Master   & Rural & 6  & Non-professional (attrition) \\
\bottomrule
\end{tabular}
\end{table}

\subsection{Procedure}
This study was conducted across five annual waves of data collection (2021–2025). To ensure longitudinal comparability, we followed the same procedure each year. Prior to participation, all participants provided written informed consent and were explicitly informed of the study’s purpose, data use, and potential implications. They were reminded of their rights to review or request deletion of their data, and to withdraw at any time without penalty.  

Each year, participants first completed a structured questionnaire (see \revise{Table~\ref{tab:questionnaire}}) covering dimensions such as cognition, usage, attitudes, and concerns, which provided quantifiable measures of group-level trends. This was followed by a 45–60 minute semi-structured interview with our researchers in which participants reflected on their experiences with generative AI over the past year, including its use in practice, evolving attitudes, and perceived impacts on their creative or professional trajectories. The full interview guide can be seen in Appendix~1. All the interviews were conducted online via Tencent Meeting and, with participant consent, were audio recorded. Recordings were transcribed into English within 48 hours and cross-checked by at least two researchers to ensure accuracy and completeness.

\revise{The survey was developed in 2021, when generative AI was still largely under-explored in the HCI community. As a result, there was limited prior work specifically examining public perceptions and attitudes toward AIGC. Nevertheless, investigating public opinions on emerging technologies is not a new research problem. Therefore, we adapted a set of theory-grounded questions from prior studies on technology perception and acceptance and contextualized them for AIGC in artistic creation. The survey primarily follows a \textit{benefits–risks evaluation framework}~\cite{bearth2016risk,tanadi2015impact, featherman2003predicting,esmaeilzadeh2020use}, which has been widely used in empirical studies of emerging technologies. First, responses to Q3–Q5 capture participants' perceived functional benefits of AIGC in the context of art creation. In particular, these questions examine perceived usefulness across different stages of the creative pipeline, including creative reference, creative tooling, and artistic expression, which represent increasing semantic levels in the artistic creation process. In contrast, Q7–Q10 probe potential risks associated with AIGC in artistic domains. These questions are informed by self-determination theory (SDT)~\cite{ryan2000self,king2007creativity}, which identifies autonomy, competence, and social context as fundamental elements of creative motivation. Because these aspects are central to artistic practice, technologies that may disrupt them are more likely to raise perceived risks. Accordingly, Q7 examines concerns about reduced creative autonomy, Q8 relates to professional competence and competitiveness, and Q9–Q10 capture broader ecosystem-level concerns such as labor market dynamics and copyright governance.  As a result, the perceived benefits and risks jointly shape individuals' overall attitudes toward the technology. These attitudes are reflected in Q11–Q12, which capture participants' general views toward AI-assisted creation. Finally, Q1–Q2 capture practitioners' behavioral outcomes, including their actual learning engagement with AIGC (i.e., Q1) and actual usage of AIGC tools in creative work (i.e., Q2) over time.}

\revise{Notably, we adopted single-item measures mainly since it is a common practice in long-term studies to maintain temporal comparability for participants, which helps sustain a reasonable retention rate. Although single-item measures could be considered less granular than multi-item constructs, prior research in psychology and organizational behavior has demonstrated their reliability and predictive validity in logitudinal studies \cite{bergkvist2015appropriate, wanous1997overall}. Moreover, single-item measures are frequently used in longitudinal designs to mitigate respondent fatigue and ensure interpretive consistency across repeated data collection waves \cite{kjaerup2021longitudinal, vogl2018developing, lewis2007analysing}. In addition, we designed the 11-point Likert scale with a range from 0 to 10, following prior longitudinal studies that adopt wider numeric ranges to provide higher sensitivity for detecting gradual attitudinal shifts over time~\cite{su2022trends, baird2010life, nebout2018comparing}. All items underwent iterative review among the research team and pilot testing with three digital painters before formal deployment to ensure clarity and linguistic neutrality.}

\begin{table}[h]
\centering
\caption{Survey Items Used in Quantitative Analysis (0–10 Likert Scale)}
\label{tab:questionnaire}
\begin{tabular}{cl}
\toprule
\textbf{Item} & \textbf{Survey Question} \\
\midrule
Q1 & I consider myself knowledgeable about the basic concepts and working principles of generative AI. \\

Q2 & I frequently use generative AI tools in my creative work. \\

Q3 & \revise{I think AI can reproduce some aspects of artistic style or expression.} \\

Q4 & \revise{I view generative AI as a tool that can assist creative work.} \\

Q5 & I think AI can provide creative references or new sources of inspiration. \\

Q6 & I am willing to spend time learning how to use generative AI effectively. \\

Q7 & \revise{Using AI may lead to homogenization of artworks and reduce individual style.} \\

Q8 & \revise{Using AI does not necessarily make artists more competitive professionally.} \\

Q9 & \revise{The widespread use of AI may reduce market demand or affect artists’ career development.} \\

Q10 & \revise{The copyright and authorship of AI-generated works are important issues that need clarification.} \\

Q11 & \revise{I have a general view, either favorable or skeptical, about the future of AI-assisted creation in digital art.} \\

Q12 & \revise{Overall, I tend to have a positive attitude toward applying generative AI to creative work.} \\
\bottomrule
\end{tabular}
\end{table}

To further guarantee data quality, all researchers underwent standardized training prior to the study to maintain consistency and depth across sessions. Research data were stored on the secure institutional cloud drive of [Anonymous University], with raw data accessible only to two researchers. All data were anonymized during analysis to protect participant privacy and confidentiality.

\subsection{Data Analysis}
Our analysis was designed to capture both longitudinal trajectories and cross-cutting dynamics in participants’ experiences with generative AI. Following guidance from longitudinal research \cite{kjaerup2021longitudinal, lewis2007analysing, vogl2018developing}, we attended to different forms of change: (a) phase shifts, where attitudes and practices moved through distinct stages across 2021–2025; (b) cross-cutting differences, such as professional status or persistent ethical concerns; (c) reinterpretations, where earlier positive or negative meanings were reframed in later years (e.g., efficiency shifting from opportunity to burden); and (d) absence of change, where stability itself—such as consistently high attention to publishing ethics—was analytically significant. Within this framework, quantitative and qualitative strands were analyzed separately and then integrated through connection and embedding.

\subsubsection{Quantitative Data Analysis}
The quantitative component was used to descriptively characterize participants’ longitudinal trajectories and to support qualitative interpretation, rather than to establish statistical generalization. Given the small sample size and exploratory nature of this study, we focused on descriptive patterns and within-participant changes over time.

After data cleaning (handling missing values and outliers), we computed means (M) and standard deviations (SD) for the \revise{twelve survey items (Q1–Q12)} at each of the five time points (2021–2025) to summarize overall trends. We also examined individual-level trajectories to identify patterns such as gradual shifts, fluctuations, and stability across waves.

Instead of inferential tests, we relied on visualizations (e.g., line plots and heatmaps) to illustrate longitudinal patterns and highlight attitudinal changes. These analyses served two purposes: (1) to provide an overview of how perceptions of generative AI evolved over time, and (2) to anchor qualitative inquiry by informing follow-up interview questions (e.g., probing reasons behind specific ratings or changes).

For comparisons between professional digital painters and hobbyists, we report descriptive differences in trajectories without making strong claims about between-group effects, given the limited and uneven sample sizes. These contrasts are primarily used to guide and contextualize qualitative analysis.

\subsubsection{Qualitative Data Analysis}
All interviews were transcribed verbatim and cross-checked for accuracy. We conducted a thematic analysis \cite{braun2006using}, supported by NVivo for data management. After getting familiar with the data, two researchers independently coded transcripts from two participants across all five waves to construct an initial codebook, which was iteratively refined with additional transcripts. Inter-coder reliability exceeded 80\%, and disagreements were resolved through discussion until consensus was reached.  

Building on this framework, we created longitudinal case files for each participant and conducted synchronous coding across their five years of data. Following recommendations for longitudinal qualitative research \cite{kjaerup2021longitudinal, lewis2007analysing, vogl2018developing}, our analysis emphasized tracing different types of change: narrative developments across waves, reinterpretations of earlier experiences, and instances of stability where accounts remained consistent over time. Finally, similar codes were aggregated into themes and subthemes, which were reviewed by the broader team to ensure coherence and distinctiveness, resulting in higher-level dynamic themes.

\subsubsection{Integration of Mixed Methods}
In the integration stage, we followed principles of ``connection'' and ``embedding'' in mixed-methods research. Connection referred to using quantitative trends to guide deeper qualitative analysis—for instance, when a survey item showed noticeable changes or fluctuations in a given year, we revisited interviews from that wave to contextualize the shift. Embedding referred to incorporating key quantitative summaries or visualizations within qualitative narratives, providing group-level references to support individual cases and making visible both convergence and divergence across methods. Rather than serving as independent evidence, quantitative results were used to complement and enrich qualitative insights, grounding interpretations in both individual experiences and overall patterns.

\section{QUANTITATIVE FINDINGS}
The quantitative results suggest that participants’ attitudes toward generative AI in painting evolved in three phases. In 2021–2022, attitudes became more negative, reaching a low point in 2022. In 2023–2024, evaluations improved across dimensions such as aesthetic value, utility, inspiration, and willingness to learn, with professional painters showing relatively higher acceptance than hobbyists. By 2025, attitudes appeared to stabilize at a generally positive level, with more modest changes over time.

\subsection{Suspicion, Awareness, and Resistance}

The years 2021–2022 represent an early stage of generative AI adoption in digital painting. During this period, participants had limited knowledge of the technology and expressed predominantly resistant attitudes. As generative AI became more visible, both awareness and experimentation increased, but negative evaluations also intensified.

\begin{figure}[htbp]
    \centering
    \includegraphics[width=0.8\textwidth]{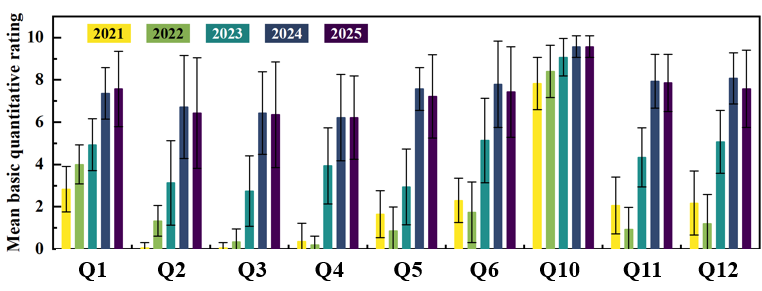}
    \caption{Mean and Standard Deviation of Participants’ Scores across Five Waves (2021--2025) on Nine Positively Scored Dimensions}
    \label{fig:quantitative_means_pos}
\end{figure}

Figure~\ref{fig:quantitative_means_pos} shows that in 2021 the mean score for \revise{Q1 (self-reported understanding of generative AI principles)} was 2.82 ($SD=1.07$), while \revise{Q2 (frequency of using generative AI in creative work)} was near zero at 0.06 ($SD=0.24$), indicating minimal familiarity and usage. As shown in Figure~\ref{fig:quantitative_scores_pos}, only P1 (a professional painter) reported any usage, while all other participants reported none. By 2022, these values increased to 4.00 ($SD=0.93$) for \revise{Q1} and 1.33 ($SD=0.72$) for \revise{Q2}, suggesting a noticeable rise in both understanding and engagement.

This pattern indicates that participants began to explore generative AI more actively and developed a clearer sense of its capabilities. Increased familiarity likely stemmed from growing exposure within creative communities rather than systematic adoption. At the same time, this early engagement coexisted with skepticism and resistance, a tension that is further unpacked in the qualitative findings (Section~5), where participants’ professional values and identity concerns played a central role.


Attitudinal responses, however, moved in the opposite direction. Scores for Q12 (overall assessment of applying generative AI to creative work) decreased from 2.18 ($SD=1.51$) in 2021 to 1.20 ($SD=1.37$) in 2022, indicating lower general endorsement. Q6 (willingness to learn to use generative AI effectively) also declined, from 2.29 ($SD=1.05$) to 1.73 ($SD=1.44$). The most pronounced decrease was observed in Q11 (general outlook on the future of AI-assisted creation in digital art), which dropped from 2.06 ($SD=1.34$) to 0.93 ($SD=1.03$). 

Together, these trends suggest that although participants became more knowledgeable and exploratory, their evaluations of AI’s potential value became more cautious. This pattern reflects a divergence between familiarity and trust, rather than outright rejection, indicating that early encounters with generative AI elicited both curiosity and concern. Such dynamics are typical of early adoption stages, where increased exposure can lead to more critical and nuanced evaluations—particularly when perceived risks to creative identity or professional stability are involved (see Section~5 for qualitative elaboration).

\begin{figure}[htbp]
    \centering
    \includegraphics[width=0.38\textwidth]{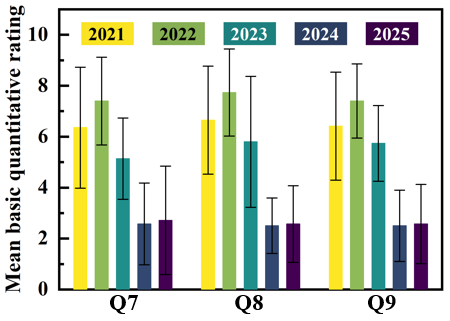}
    \caption{Mean and Standard Deviation of Participants’ Scores across Five Waves (2021--2025) on Three Reverse-Scored Dimensions}
    \label{fig:quantitative_means_neg}
\end{figure}

\begin{figure}[htbp]
    \centering
    \includegraphics[width=0.55\textwidth]{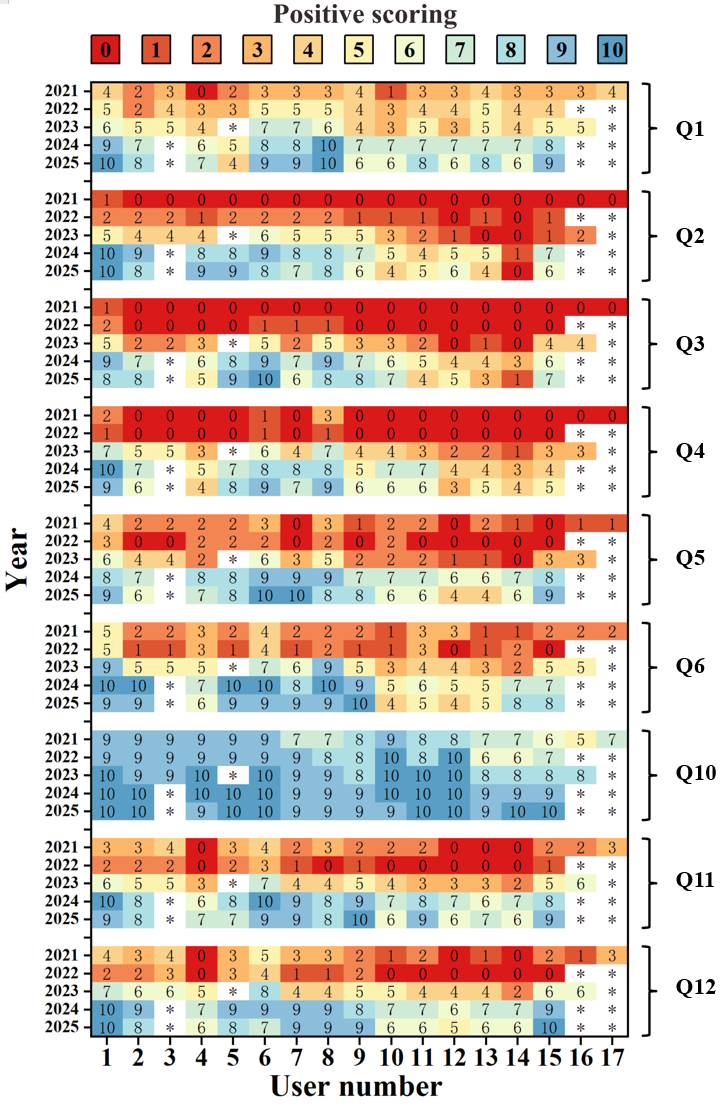}
    \caption{Participants’ Scores across Five Waves (2021--2025) on Nine Positively Scored Dimensions}
    \label{fig:quantitative_scores_pos}
\end{figure}

Responses to items concerning potential risks showed a similar pattern of heightened concern (Figure~\ref{fig:quantitative_means_neg}), with detailed individual-level patterns shown in Figure 4. Scores for Q8 (belief about AI’s influence on professional competitiveness) increased from 6.65 ($SD=2.12$) in 2021 to 7.73 ($SD=1.71$) in 2022, suggesting growing awareness of competitive pressures within the art community. Other items, such as Q7 (concern about homogenization and loss of individual style when using AI) and Q9 (concern about reduced market demand due to AI adoption), also showed upward trends. 

Overall, these descriptive patterns characterize this phase as one of skepticism and caution rather than outright rejection. Participants began to engage with generative AI while simultaneously expressing concerns about its creative and economic implications—an ambivalence that is further explored in the qualitative findings.

\subsection{Rapid Acceptance of Generative AI}

Between 2022 and 2024, participants’ responses to Q12 (overall assessment of applying generative AI to creative work) improved markedly. Scores increased from 1.20 ($SD=1.37$) in 2022 to 5.07 ($SD=1.49$) in 2023, and further to 8.07 ($SD=1.21$) in 2024, indicating a clear positive shift in overall evaluation. This pattern suggests that early skepticism gave way to growing acceptance as generative AI became more visible and practically useful within participants’ creative routines.

As shown in Figure~\ref{fig:quantitative_means_pos}, positive changes extended across multiple items. \revise{Q3 (perception of AI’s ability to reproduce artistic style or expression)} increased from 0.33 ($SD=0.62$) in 2022 to 2.73 ($SD=1.67$) in 2023, and to 6.43 ($SD=1.95$) in 2024, reflecting a substantial shift in perceived capability. Similar upward trends were observed in \revise{Q4 (AI as an auxiliary creative tool)}, Q5 (AI as a source of inspiration), Q6 (willingness to learn to use generative AI effectively), and Q11 (general outlook on AI-assisted creation). Notably, increases in inspiration (Q5) and learning willingness (Q6) were particularly pronounced between 2023 and 2024, suggesting deeper engagement beyond initial exploration.

Together, these patterns indicate that participants increasingly recognized AI’s creative affordances and became more motivated to integrate it into their workflows through active experimentation and learning.

In parallel, \revise{responses to items reflecting perceived risks} declined substantially (Figure~\ref{fig:quantitative_means_neg}). \revise{Q7 (concern about homogenization and loss of individual style)} decreased from 7.40 ($SD=1.72$) in 2022 to 5.13 ($SD=1.60$) in 2023, and further to 2.57 ($SD=1.60$) in 2024. Comparable downward trends were observed for \revise{Q8 (concern about professional competitiveness)} and \revise{Q9 (concern about reduced market demand)}. 

These changes suggest that as familiarity and confidence increased, concerns about competition and creative loss became less dominant, reflecting a gradual normalization of generative AI within participants’ professional environments.

\revise{Overall, the 2022–2024 period represents a rapid adoption phase characterized by substantial attitudinal improvement and expanding engagement. While these quantitative patterns illustrate broad shifts, the qualitative analysis (Section~5) provides deeper insight into how evolving perceptions of work, creative value, and professional identity shaped these changes over time.}

\begin{figure}[htbp]
    \centering
    \includegraphics[width=0.55\textwidth]{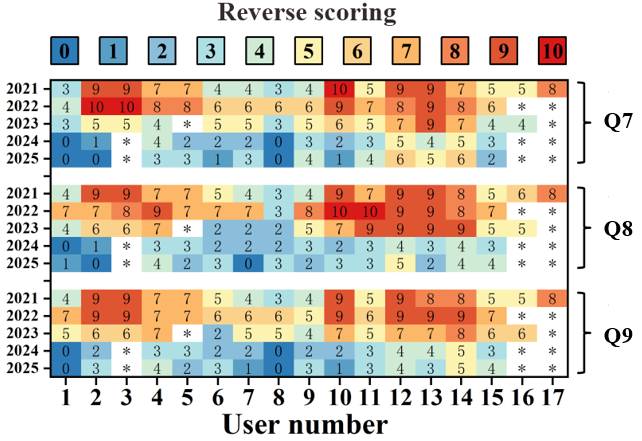}
    \caption{Participants’ Quantitative Scores across Five Waves (2021--2025) on Three Reverse-Scored Dimensions}
    \label{fig:quantitative_scores_neg}
\end{figure}

\subsection{Stabilization Phase}

From 2024 to 2025, participants’ \revise{responses to Q12 (overall assessment of applying generative AI to creative work)} remained largely stable, with a slight decrease from 8.07 ($SD=1.21$) to 7.57 ($SD=1.83$). Similar mild declines were observed in \revise{Q3 (perception of AI’s ability to reproduce artistic style or expression)} and \revise{Q6 (willingness to learn to use generative AI effectively)}. In contrast, \revise{Q1 (self-reported understanding of generative AI principles)} showed a small increase from 7.36 ($SD=1.22$) to 7.57 ($SD=1.79$). \revise{Q10 (belief in the importance of clarifying copyright and authorship)} remained consistently high ($M=9.57$, $SD=0.51$). As shown in Figure~\ref{fig:quantitative_means_neg}, \revise{items addressing perceived risks (Q7–Q9)} also remained relatively stable over this period.

\revise{This period can be interpreted as a stabilization stage following the rapid adoption observed in earlier years. Overall attitudes remained highly positive, with only minor fluctuations, suggesting that initial enthusiasm had reached a plateau. Rather than continued growth, the dominant pattern was consolidation: participants appeared to have integrated generative AI into their everyday workflows, and their evaluations became more consistent and pragmatic. These patterns indicate a shift from novelty-driven engagement toward sustained and normalized use. The qualitative analysis (Section~5) further examines how this stabilization intersected with evolving creative practices and ethical reflections.}

\subsection{Attention to AI Ethics}

Unlike the dynamic changes observed in other dimensions, participants’ \revise{responses to Q10 (belief in the importance of clarifying the copyright and authorship of AI-generated works)} remained consistently high and increased over time. Scores rose from 7.82 ($SD=1.24$) in 2021 to 8.40 ($SD=1.24$) in 2022, 9.07 ($SD=0.88$) in 2023, and reached 9.57 ($SD=0.51$) in 2024, remaining at this level in 2025.

\revise{These patterns indicate that ethical considerations around copyright and authorship remained highly salient throughout the five-year period. While perceptions of AI’s creative utility evolved across phases, consistently high ratings on this item suggest that ethical awareness became a stable and integral component of participants’ overall understanding of generative AI.}

\subsection{Professional Status and Attitudinal Differences}

\begin{figure}[htbp]
    \centering
    \includegraphics[width=0.5\textwidth]{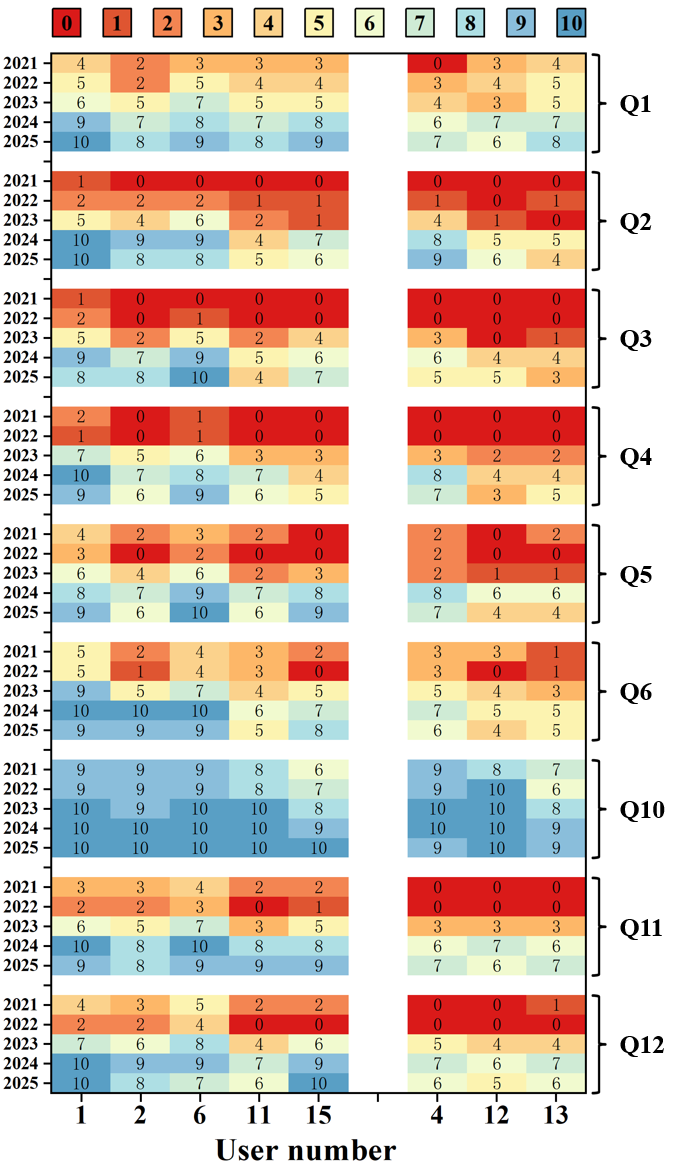}
    \caption{Attitudes of Professional (left) and Non-professional Artists (right) toward AI-assisted drawing on positively scored items.}
    \label{fig:attitudes_positive}
\end{figure}

\begin{figure}[htbp]
    \centering
    \includegraphics[width=0.5\textwidth]{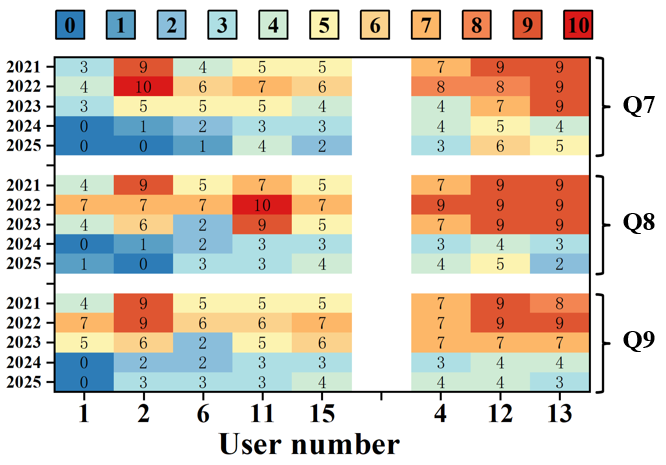}
    \caption{Attitudes of Professional (left) and Non-professional Artists (right) toward AI-assisted drawing on negatively scored items.}
    \label{fig:attitudes_negative}
\end{figure}

\begin{figure}[htbp]
    \centering

    \includegraphics[width=0.7\textwidth]{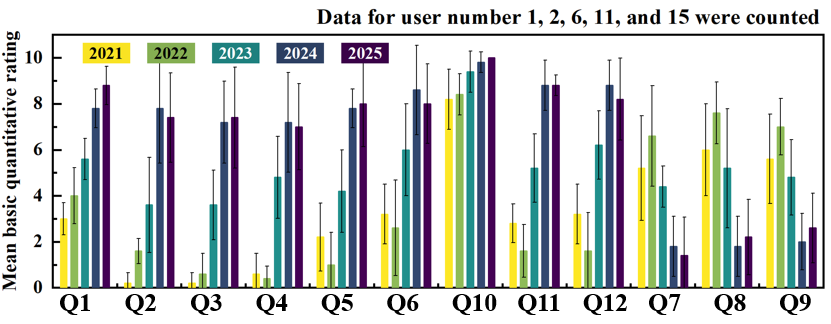}

    \vspace{1em}
    \includegraphics[width=0.7\textwidth]{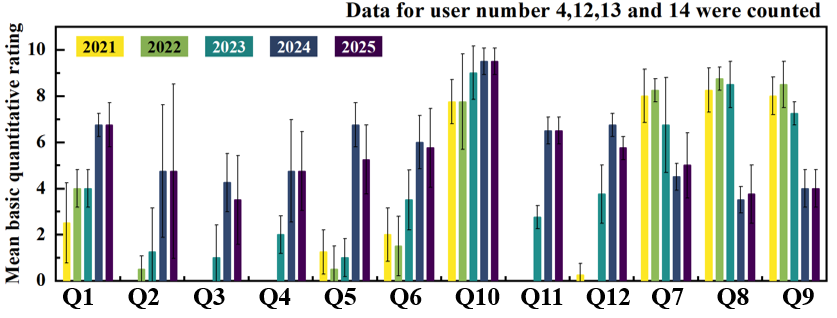}

    \caption{Mean quantitative ratings of Professional Artists (top, Users 1, 2, 6, 11, and 15) and Non-professional Artists (bottom, Users 4, 12, 13, and 14) across 2021--2025.}
    \label{fig:quant_professional_vs_nonprofessional}
\end{figure}

As shown in Figures~\ref{fig:attitudes_positive}–\ref{fig:quant_professional_vs_nonprofessional}, professional digital painters and hobbyists exhibited systematic differences in their attitudes toward generative AI. Overall, professionals showed higher acceptance levels and more positive trajectories over time, while hobbyists demonstrated more moderate improvements.

\revise{To facilitate a consistent comparison between groups, Figures~\ref{fig:attitudes_positive}–\ref{fig:quant_professional_vs_nonprofessional} visualize a subset of participants ($N=8$) selected based on longitudinal completeness and stable professional status. Participants with incomplete data due to attrition (e.g., P3, P17) or changes in professional identity (e.g., P10) were excluded. The final sample includes individuals who maintained consistent roles and provided complete data across all five waves (Professionals: P1, P2, P6, P11, P15; Non-professionals: P4, P12, P13). While limited in size, this subset enables a clearer descriptive comparison between participants embedded in professional production contexts and those engaging in creative work as a personal pursuit.}

Across the nine positively scored dimensions (Figure~\ref{fig:attitudes_positive}), professional painters consistently reported higher ratings than hobbyists from 2021 to 2025, with the gap becoming more visible in later years. For example, on \revise{Q12 (overall assessment of applying generative AI to creative work)}, professionals reached an average score close to 8 by 2025, compared to around 6 among hobbyists. Similar patterns were observed in \revise{Q5 (AI as a source of inspiration)} and \revise{Q6 (willingness to learn to use AI effectively)}, where professionals showed stronger and more sustained engagement.

Conversely, across the negatively scored dimensions (Figure~\ref{fig:attitudes_negative}), professionals generally reported lower levels of concern than hobbyists, suggesting a comparatively lower perception of threat or disruption. For instance, by 2025, the mean score for \revise{Q9 (concern about reduced market demand)} was approximately 2.5 among professionals, compared to nearly 4 among hobbyists.

\revise{These descriptive contrasts suggest that professional painters may have been more adaptive and confident in engaging with generative AI, whereas hobbyists tended to remain more cautious, expressing greater uncertainty and concern about potential impacts. At the same time, given the small and uneven group sizes, these differences should be interpreted as indicative patterns rather than definitive group effects. While the numerical trends highlight professional status as an important contextual factor, they do not, on their own, explain the underlying mechanisms. The qualitative analysis (Section~5) further examines how factors such as economic dependence, workflow integration, and community influence contributed to these divergent trajectories.}

\section{QUALITATIVE FINDINGS}
The qualitative findings trace how painters’ experiences with generative AI evolved over time: from early resistance (2021–2022), to pragmatic adoption (2022–2024), and finally to reflective reconsideration (2025). In addition to this longitudinal trajectory, cross-cutting themes of peer influence and emotional shifts highlight the social and affective dynamics that shaped adoption, complementing the quantitative trends with deeper explanations of why attitudes changed.

\subsection{Longitudinal Trajectory of Attitudes}

\subsubsection{Strong Resistance}

Between 2021 and 2022, participants moved from tentative curiosity to outright resistance toward generative AI in digital painting. In 2021, many described themselves as “testing the waters”—interested but uncertain. Several downloaded free AI apps or tried early demo websites, but treated them more as novelties than serious tools. As P3 recalled, \textit{“I played with it for fun, just to see what it could do. At that time it felt like a toy, not something that could threaten me” (2021).} By 2022, however, as AI-generated images became widely circulated on social media and art forums, the tone shifted dramatically. What had seemed like harmless experimentation quickly turned into disillusionment and even hostility. Participants frequently dismissed the outputs as “garbage,” “worthless collages,” or \revise{“devoid of artistic sensibility.”} P1, who in 2021 had spoken of curiosity, now concluded: \textit{“I tried again this year, and honestly, it became worse. Everyone is posting AI art everywhere, and to me they all look the same—shallow, soulless” (2022).} This illustrates a turning point where initial exploration hardened into distrust once participants confronted AI’s growing visibility in their creative environments.

\revise{The most common criticism concerned the perceived artistic shallowness of AI-generated images.} In 2021, some still wondered whether AI images might one day achieve artistic merit. But by 2022, nearly all rejected them as inauthentic. P4 gestured to several online images and remarked, \textit{“Look at these things. Do you think they have any value? They are nothing but visual gibberish” (2022).} P9 added, \textit{“Maybe they look polished, but they have no warmth. Real art always carries traces of the artist’s struggle, and these images erase that struggle” (2022).} Such comments reveal not just dissatisfaction with technical quality but a deeper anxiety about whether art without human touch can still be called art.

A second recurring critique targeted creativity itself. Ten participants emphasized that creativity is a uniquely human process that cannot be reduced to recombination. As P12 argued, \textit{``Can patching things together really count as creativity? If so, the bar for creativity is far too low'' (2022).} \revise{A closely related concern involved the perceived erosion of human authorship and artistic identity.} Several participants stressed that true painting requires human observation and decision-making, whereas AI generation bypasses this process and thus cannot be considered authentic creation. P5 elaborated:  
\begin{quote}
    ``Art is fundamentally a human act of creation—we observe the world, choose our tools, sketch with pencils, color with markers, each step full of human uncertainty. AI’s collages skip this process entirely, and therefore cannot be seen as creation'' - P5, (2022).
\end{quote} 

Others worried about \revise{loss of personal recognition and artistic credit.} P14 asked, \textit{``If people can’t tell the difference, what does my signature mean anymore? My name used to guarantee originality. Now it might not'' (2022).} \revise{This perceived blurring of authorship boundaries underscored how generative AI was seen as undermining the symbolic value of artistic ownership.}  

Finally, \revise{copyright and data-use anxieties further intensified participants’ distrust.} Some feared their works could be scraped into training datasets without consent. P7 worried, \textit{``There are no proper laws about AI right now. If someone takes my work to train a dataset, do I still own the outputs? Could my work be misused by others?'' (2022).} For professionals especially, this was tied to livelihood. P2 put it bluntly: \textit{``If AI can borrow from my drawings without asking, it’s not just unfair—it’s stealing my future opportunities'' (2022).}

In conclusion, the early stage of 2021–2022 marked a trajectory from hesitant exploration to entrenched resistance. While quantitative results showed that attitudes reached their lowest point in 2022, the qualitative narratives explain why: participants saw AI images as aesthetically hollow, rejected their creative legitimacy, and feared both authorship erosion and legal uncertainty. This phase underscores a broader tension in human–AI interaction: even as familiarity grows, acceptance cannot follow when users perceive a technology as undermining the very values that define their professional and creative identity.

\subsubsection{Instrumental and Pragmatic Adoption}
Between 2022 and 2024, participants’ attitudes toward AI-assisted painting shifted dramatically. What had been framed as an unworthy imitation in 2022 gradually became recognized as a practical tool. This trajectory was neither linear nor uniform: some participants approached AI with deliberate curiosity, while others adopted it reluctantly under the pressure of peers, clients, and industry competition. The change aligns with quantitative patterns showing steep rises across multiple positive dimensions (see Section~5.1.2). Qualitative analysis further reveals the mechanisms behind this shift. In 2022, transcripts were filled with negative descriptors such as ``ugly,'' ``trash,'' and \revise{``devoid of artistic appeal,''} appearing 155 times. By 2023 these had dropped to 34 mentions, and by 2024 they had nearly disappeared—replaced by evaluations that emphasized \revise{practical utility, visual quality, and creative usefulness.} This linguistic shift captures the broader move from skepticism to conditional acceptance.  

\textbf{\revise{Reappraising artistic quality.}} By 2023, improvements in underlying models and training data were widely noticed. Several participants admitted that, while they still harbored doubts about creativity, they could not deny the visual appeal of many outputs. P6 reflected on this shift: \textit{``I saved many AI-generated comic characters on my phone, not for any particular reason other than that their color schemes look really appealing, and I can use them as references when drawing'' (2023).} Similarly, P11 acknowledged the increasing polish: \textit{``If you don’t explicitly disclose it, many AI artworks are already hard to distinguish from human-made pieces, and in some cases they even surpass the quality of ordinary painters'' (2023).} What had once been dismissed as ``garbage'' was now quietly folded into reference collections and mood boards. \revise{This collective reassessment of visual quality mirrored the broader trend of acceptance observed in survey data, suggesting that digital painters began to perceive AI images not as replacements for creativity but as valuable visual resources.}

\textbf{\revise{Efficiency as adaptation.}} For professionals, the more decisive factor was not aesthetics but survival. By 2023–2024, many described AI as unavoidable for commercial workflows, even if they disliked it. P10 captured this ambivalence: \textit{``Even if I don’t like AI art, I have no choice—my colleagues are already using it. What takes me ten days to draw, AI can finish in a minute. I still stubbornly believe my work looks better, but that doesn’t matter'' (2024).} P2 echoed this pragmatic tone: \textit{``Using AI to generate background images is extremely convenient—feed it a dataset and it’s done in a few minutes'' (2024).} \revise{These accounts illustrate how AI was no longer regarded as a peripheral novelty but had become embedded in production practices, valued for its capacity to compress labor and time costs, even when doubts about artistic integrity persisted.}

\textbf{\revise{AI as an ideation scaffold.}} Beyond speed, participants began to incorporate AI into early stages of conceptualization and communication. Rapid iteration enabled them to explore directions and interact with clients more effectively. P15 described this workflow: \textit{``Many times I just tell the AI the general idea, let it generate a rough sketch, and then use that sketch to communicate with the client—so I don’t need to draw it myself'' (2024).} \revise{Others described AI as a means of prototyping visual possibilities or triggering inspiration, treating the generated results as raw material rather than finished work.} By reframing AI as a \revise{supporting collaborator in idea development}, painters shifted from resisting its presence to positioning it as a facilitator within the broader creative process.  

\revise{In summary, between 2022 and 2024, digital painters moved from suspicion to strategic incorporation of AI within their creative workflows. Although their motivations varied—from efficiency to experimentation—the underlying logic was pragmatic: AI was integrated not as a symbol of artistry but as a versatile instrument for managing time, cost, and creative uncertainty. This pragmatic adoption phase parallels the positive trajectory observed in survey trends, representing the moment when generative AI transitioned from an external threat to an operational ally in artistic production.}

\subsubsection{Identity Reconstruction and Value Reconsideration}
By 2025, participants’ attitudes toward generative AI took a new turn. Compared to the heightened acceptance observed in 2024, enthusiasm had cooled and critical voices resurfaced. Yet the focus of critique had shifted: rather than debating whether AI images looked good enough, participants now grappled with what AI meant for their livelihoods, their artistic values, and their very sense of identity. \revise{This reflective stage corresponded with the overall stabilization seen in survey trends (see Section~5.1.3), suggesting that initial excitement had given way to more nuanced and ambivalent evaluations.}  

\textbf{\revise{Labor and industry pressures.}} For many professionals, efficiency gains no longer inspired optimism but instead revealed uncomfortable truths about labor and compensation. Five participants reported shrinking salaries, reduced job opportunities, or even unemployment. P8 observed, \textit{``If AI can finish in three minutes what I do in a month, of course my boss won’t pay me 7,000 RMB anymore. That’s just normal’’ (2025).} P2 noted the irony of increased productivity without reward:  
\begin{quote}
    ``In the past, I produced one game concept drawing per week and was already one of the fastest in my company. Now, with AI assistance, I can make three per day, but my salary has barely changed. It’s ridiculous.’’ - P2, (2025).
\end{quote} 
\revise{Others voiced deeper concerns about professional meaning and self-worth. P4 confessed, \textit{``I have no meaning anymore—it (AI) can do everything’’ (2025).} What had once been celebrated as efficiency in 2023–2024 was reinterpreted as precarity in 2025, revealing how technological acceleration reshaped creative labor hierarchies and intensified feelings of replaceability.}

\textbf{\revise{Aesthetic fatigue and loss of texture.}} Six participants reported that AI’s technical polish had paradoxically produced monotony. While in 2023 many admired its rapid improvements, by 2025 some felt uneasy with its ``too perfect’’ qualities. As P13 explained:  
\begin{quote}
    ``Although today’s AI drawings are no longer obviously AI, if you look closely you can still tell. There are no traces of handcraft—too perfect in a way that feels unsettling.’’ - P13, (2025).
\end{quote}
\revise{Several others echoed this sentiment, describing AI-generated imagery as visually impressive but emotionally hollow. The same precision that once symbolized progress now evoked fatigue, prompting a renewed longing for imperfection and human touch.}

P9 echoed this sense of saturation: \textit{``Everywhere I look—platforms, forums, galleries—it’s flooded with AI works. After a while, they all start to look the same’’ (2025).} \revise{Such reflections indicate that under the flood of highly similar and mass-produced AI imagery, painters were no longer only worried about displacement but also began to question the cultural depth and diversity of AI outputs.}

\textbf{\revise{Reclaiming identity and authorship.}} Most strikingly, participants began to renegotiate their professional identities. \revise{Some emphasized human-made work as an ethical and aesthetic stance, drawing clearer boundaries between human and algorithmic contributions.} Two explicitly stated they would limit AI to early ideation or reference stages to avoid ``AI-ization’’ of their work. Others sought hybrid workflows that preserved a sense of authorship. P14 described his strategy in detail:  
\begin{quote}
    ``My work has four steps: hand-drawn sketch, AI coloring, manual correction, and fine-tuning. This way I can still guarantee the quality of the drawing throughout the process, and my work feels meaningful.’’ - P14, (2025).
\end{quote} 
P6 offered a more symbolic view: \textit{``Even if I use AI halfway, I always leave some parts purely mine—like the character’s face. That’s where I feel my identity still lives’’ (2025).} \revise{These accounts reveal a transition from passive adaptation to active negotiation, in which digital painters redefined creative control and reasserted personal meaning within hybrid workflows. Rather than rejecting AI outright, they sought to integrate it selectively to maintain a sense of integrity and self-expression.}

\revise{Overall, the narratives from 2025 portray a period of reflection rather than celebration. After the phase of rapid adoption, participants approached generative AI with a more measured and differentiated stance. They recognized that AI had become deeply embedded in their routines yet simultaneously questioned its implications for artistic value and human distinctiveness. This reflective equilibrium marks a shift from fascination toward critical coexistence, where digital painters reconsidered what creativity, labor, and authorship mean in an AI-saturated environment.}

\subsection{Cross-Cutting Themes}
\subsubsection{Peer Influence and Social Negotiation}

Beyond individual attitudes, \revise{peer and community dynamics played a decisive role in shaping how painters interpreted and engaged with generative AI. Participants consistently located their own decisions within the expectations, habits, and discourses of their professional and social circles, rather than viewing them as purely personal choices.} This theme became especially salient between 2022 and 2025, \revise{coinciding with the broader diffusion of AI practices documented in survey trends} (see Section~5.1.2).  

In 2022, peer influence was often experienced as pressure rather than encouragement. Several participants described feeling obliged to try AI simply because colleagues or friends were experimenting with it. P10 explained, \textit{``Even if I didn’t want to use it, my colleagues kept sending AI drafts in our group chat. If I ignored them, I felt out of touch’’ (2022).} P3 echoed this tension: \textit{``I didn’t trust it, but everyone around me was testing it. I was worried I’d be left behind if I didn’t at least try’’ (2022).} \revise{These accounts indicate that early exploration was often motivated by social alignment rather than intrinsic curiosity, showing how creative communities functioned as informal agents of diffusion.}  

By 2023–2024, peer influence \revise{evolved from subtle pressure to social normalization}. Participants recalled AI-generated images circulating widely across forums, WeChat groups, and online galleries, where sharing outputs became routine. P11 observed, \textit{``Last year people laughed at AI drawings. Now, in our group, everyone posts them casually, like another draft. Nobody finds it strange anymore’’ (2023).} Similarly, P7 noted how client expectations had adjusted: \textit{``If you don’t use AI for concept sketches, clients may even ask why. They assume everyone is using it now’’ (2024).} \revise{These examples show how collective habits, visibility, and client norms transformed generative AI from a peripheral novelty into a socially expected practice, aligning with—but not reducible to—the quantitative patterns of broader acceptance.}  

By 2025, however, \revise{community negotiations had become more contested}. Some participants described distancing themselves from AI-heavy groups to preserve artistic integrity, while others doubled down on hybrid workflows and defended AI as a legitimate professional tool. P14 captured this polarization: \textit{``Some of my friends insist that real artists must reject AI completely, while others post AI drafts every day. Our conversations sometimes turn into arguments—I feel caught in between’’ (2025).} \revise{These tensions suggest that even after normalization, the social life of AI remained unsettled—its meanings continually debated and reinterpreted within peer discourse.}  

\revise{In Conclusion, these narratives reveal that adoption was not an isolated psychological progression but a socially situated negotiation. Peer dynamics simultaneously accelerated experimentation and later provoked moral and professional reflection. This underscores that in creative domains, trajectories of human–AI interaction are co-constructed through collective sensemaking rather than shaped solely by individual learning or preference.}

\subsubsection{Emotional and Experiential Shifts}

Alongside changing attitudes, participants’ emotional responses to generative AI evolved markedly over the five years, \revise{revealing how affective experience intertwined with technological familiarity and creative adaptation. These shifting emotions illuminate not only what participants thought about AI, but how they felt while negotiating its presence in their artistic lives.}  

In 2021, curiosity dominated. Early encounters were described as playful experiments, often tinged with humor. P3 recalled, \textit{``At first I treated it like a toy. I typed in silly prompts just to see what came out, and it was funny more than anything else’’ (2021).} \revise{This lighthearted curiosity represented a low-stakes mode of exploration, where AI was still framed as a novelty rather than a serious creative partner.}  

\revise{As the technology gained visibility in 2022, however, affect shifted sharply toward frustration and anger.} Participants described feeling disillusioned and even offended by what they perceived as low-quality imitations of human art. P4 exclaimed, \textit{``It’s everywhere now, and it makes me mad. These so-called artworks look the same and clutter the platforms’’ (2022).} P12 echoed this hostility: \textit{``When people call these collages ‘creative,’ it feels like an insult to years of training’’ (2022).} \revise{Such responses reveal how emotional resistance was rooted less in aesthetic disagreement than in perceived threats to effort, skill, and artistic legitimacy.}  

By 2023–2024, emotions shifted toward excitement and even awe. Participants who had once mocked AI now expressed surprise at its progress. P6 admitted, \textit{``I was shocked—the images suddenly looked good. Sometimes I even felt inspired, like it gave me new ideas’’ (2023).} P11 described a moment of exhilaration: \textit{``The first time I saw an AI piece that truly amazed me, I felt both scared and thrilled. It was better than many human works’’ (2024).} \revise{This ambivalence of fear and fascination—what some described as being both impressed and unsettled—reflected the emotional novelty effect that often accompanies rapid technological breakthroughs.}  

By 2025, however, the emotional tone had cooled. Participants spoke of fatigue, boredom, and unease with the overabundance of AI images. P13 explained, \textit{``I don’t feel shocked anymore. AI works are everywhere, and they all feel a bit too perfect—after a while it’s numbing’’ (2025).} P9 noted the loss of personal attachment: \textit{``When I scroll through feeds full of AI art, I feel indifferent. Nothing moves me anymore’’ (2025).} For some, this fatigue opened space for reflection. As P14 summarized: \textit{``At first I was excited, then I felt overwhelmed, and now I just want to rethink what role it should play in my work’’ (2025).} \revise{These narratives show that emotional desensitization did not simply signal withdrawal but the beginning of a more reflective relationship with technology.}  

\revise{In sum, these affective trajectories—from playful curiosity to frustration, from excitement to reflective fatigue—illustrate that engagement with AI was as much an emotional negotiation as a cognitive or instrumental one. Rather than treating emotional change as a byproduct of adoption, these experiences suggest that long-term integration of AI depends on how technologies sustain curiosity, manage ambivalence, and support users’ evolving sense of creative connection.}

\section{DISCUSSION}
Our five-year longitudinal study reveals how Chinese digital painters' engagement with generative AI evolved—shaped not only by shifting attitudes over time, but also by peer dynamics and emotional experience. Rather than simple tool adoption, their journey reflects a deeper, social negotiation of identity and values. We discuss in  (1) longitudinal trajectories, (2) lateral influences, (3) human-AI design implications, and (4) limitations and future work.

\subsection{Longitudinal Trajectories of Attitude and Identity Negotiation}
Our longitudinal findings reveal that digital painters’ attitudes toward generative AI did not follow a linear path of acceptance, but instead evolved through distinct phases: early resistance and skepticism, mid-stage pragmatic adoption, and late-stage critical reflection and identity reconstruction. This dynamic trajectory complements existing adoption research by addressing the limitations of “gradual acceptance” or “rejection–acceptance binary” models \cite{rogers2014diffusion, davis1989technology,sikorski2025attitudes, johnston2024understanding}, and suggests that creators’ relationship with AI is better understood as a process of ongoing negotiation rather than a one-dimensional adoption curve.

In the early stage (2021–2022), resistance was driven not only by disappointment with the quality and stability of outputs, but also by deeper ontological doubts about the nature of art. Many participants maintained that art is a uniquely human form of emotional and creative expression, whereas AI merely recombines patterns based on data. This form of "ontological skepticism" led them to perceive AI as a threat, particularly due to concerns that their style and labor might be learned or imitated without consent \cite{kirova2023ethics, lima2025public}. These anxieties were not isolated but intertwined with public imaginaries, ethical controversies, and copyright fears circulating at the time—intensifying their resistance.

As generative AI tools significantly improved during 2023–2024, attitudes shifted toward a more utilitarian mode of acceptance. Especially among professional painters, adoption was shaped by industry pressures and economic demands: AI was reframed as a tool for efficiency. Many emphasized that refusing to use AI meant losing competitiveness or failing to meet client expectations, prompting them to learn and integrate AI features pragmatically. This “instrumental turn” closely mirrors how creative professionals in other domains have adapted to emerging technologies \cite{page2025creative, chung2022artistic, canet2022dream}. In contrast, hobbyists’ changes were more dependent on personal interests or creative philosophy; lacking the same occupational risks, they often remained hesitant or resistant for longer. This contrast underscores the value of a longitudinal lens, which reveals how differences in professional status lead to divergent adoption paths shaped by institutional and economic contexts \cite{tang2024exploring, inie2023designing}.

By 2025, a new layer of complexity emerged. Although earlier resistance had faded, the initial excitement had also cooled—giving way to reflection and identity negotiation. Some artists explicitly constrained the use of AI (e.g., limited to ideation or rough sketching), while others deeply integrated AI while retaining creative control. Still others explored new human–AI hybrid modes, treating AI as an extension of their creative agency. These diverse practices suggest a shift from asking “whether to use AI” toward the more nuanced question of “how to maintain distinctiveness while using AI.” This process of identity negotiation highlights creators’ agency in the face of technological change—they actively reconstructed the role of AI in their workflow to preserve their own value and professional identity \cite{munoz2023identity, rezwana2025human, yang2020re, moruzzi2024user}.

A longitudinal perspective allows us to see that these phase-based shifts are not simply reactions to technological progress, but part of an ongoing “negotiation of identity and value” \cite{kjaerup2021longitudinal, saldana2003longitudinal}. Without such a view, cross-sectional studies might misinterpret resistance as “backwardness,” or acceptance as a final “endpoint,” missing the evolving relationship between creators and technology. The contribution of this study lies in revealing this trajectory and complexity, offering a more nuanced understanding that AI technology adoption is not merely about utility, but about the ongoing reconstruction of identity, value, and social relationships.

\subsection{Lateral Dynamics of Peer Influence and Emotional Trajectories}
Beyond the three-phase trajectory of adoption, our findings highlight two lateral forces that shaped creators’ engagement with generative AI: peer influence and emotional experience. These forces intersected with individual attitudes over time, revealing that adoption is not only a function of tool capabilities, but also of social relationships and affective contexts \cite{bran2023emerging}.

Peer influence played a sustained role throughout the five-year period. While many initially resisted AI, they began adopting it as peers embraced it, clients demanded it, and communities normalized its use. This was not simple imitation, but a form of normative pressure shaped by professional ecosystems: rejecting AI risked marginalization or loss of relevance \cite{vannoy2010social, woodruff2024knowledge, zhang2024confrontation}. At the same time, peer dynamics amplified divergence. Some gained early advantage and pioneered new modes of collaboration, while others doubled down on rejection, viewing AI adoption as a betrayal of artistic ethics. These tensions underscore that peers act not only as conduits for diffusion but also as agents of polarization within creative communities.

Emotional experience also shaped how creators responded to AI. Participants described a shifting emotional arc—from early anticipation and anxiety, to mid-stage excitement, and later fatigue and ambivalence \cite{cetinic2022understanding, johnston2024understanding}. Initial anxiety was tied to defensiveness and copyright fears; mid-phase enthusiasm emerged from productivity gains and novelty; later fatigue stemmed from aesthetic saturation and rising cognitive load. These emotions were not incidental but actively influenced the nature and depth of engagement.

Importantly, peer influence and emotional dynamics were mutually reinforcing. Broad adoption heightened fears of falling behind; shared enthusiasm fostered collective excitement and, at times, hype. Conversely, as communities fragmented, emotions shifted toward fatigue or reflective disengagement. This interplay—what we term \textit{social-emotional resonance}—helps explain the variability and volatility of individual adoption paths \cite{lang2025peer, wang2025exploring}.

Above all, these insights call for a broader understanding of creativity support systems. Prior design work often centers on user–tool interaction, overlooking how group norms and emotional climates shape use \cite{bennett2024painting, lu2025designing, inie2023designing, hu2025designing}. We argue that designers must recognize the social ecology surrounding creative tools. Systems should accommodate individual rhythms and offer flexible boundaries to ease peer-induced pressure. They should also support emotional self-regulation, helping users channel anxiety or fatigue into sustainable, meaningful engagement \cite{slovak2023designing, liu2025regulation}.

\subsection{Design and Theoretical Implications for Human--AI Collaboration}
Our study offers a rare longitudinal perspective on how generative AI reshapes creative practices over time. While prior HCI work often relies on cross-sectional surveys or short-term experiments \cite{lovato2024foregrounding, deshpande2024perceptions, ma2024drawing}, our five-year investigation reveals that creators’ relationships with AI evolve through sustained interactions—shaped not only by technical affordances, but also by shifting social dynamics and ongoing identity negotiation. This view challenges design approaches based solely on early user feedback, underscoring the need to consider how human–AI relations are continuously constructed and redefined.

A core design implication centers on agency and control. While participants welcomed AI for ideation, exploration, and localized support, they resisted tools that attempted to dominate the creative process. This resistance reflects creators’ desire to maintain authorship and influence. Thus, systems should avoid full automation as a goal. Instead, they should offer adjustable parameters, interpretable outputs, and granular controls—positioning AI as a flexible medium rather than an opaque agent.

Our findings also reveal a wide spectrum of human–AI collaboration strategies. Some creators use AI only for early-stage inspiration; others embed it throughout their workflow while retaining oversight; still others experiment with co-creative paradigms. This diversity suggests that no single interaction model suffices. Future systems should support layered collaboration—from lightweight assistance to immersive integration—and allow fluid transitions across modes. Such flexibility respects creators’ evolving goals and differing relationships with AI.

Notably, many participants embraced failure as a creative resource. Imperfect or unexpected outputs—glitches, inconsistencies, stylistic clashes—often sparked new directions. Rather than correcting all errors, creators preferred tools that preserved ambiguity and invited reinterpretation. This insight aligns with prior HCI work on serendipity and productive failure \cite{yang2020re, lu2025designing, park2024we, he2023exploring}. Designers should consider enabling “controlled failure,” where imperfection is not a flaw to be fixed but a spark for exploration.

On a theoretical level, our results highlight how generative AI prompts a dynamic renegotiation of creative identity. Over time, participants moved from essentialist skepticism (“AI undermines human creativity”) to instrumental pragmatism (“AI helps meet deadlines”) to reflective rebalancing (“AI is a collaborator, not a replacement”). This trajectory demonstrates that adoption is not merely functional—it is deeply intertwined with value systems and professional positioning.

Finally, our work expands the theoretical framing of creative support tools by incorporating emotion and social dynamics. Prior HCI work has often emphasized usability, functionality, and performance, while overlooking the affective trajectories and peer pressures that accompany long-term tool use \cite{rezwana2025human, xie2025embodied, rezwana2025human, kyi2025governance}. Our data show how excitement, anxiety, fatigue, and normative expectations co-evolve with users’ tool preferences. These insights point to a more socially and emotionally grounded understanding of creative interaction, reminding designers to attend not only to human–machine interfaces, but to the broader social ecologies in which they are embedded.

\subsection{Limitations and Future Work}
While this study provides a rare five-year longitudinal lens on creators' evolving relationships with generative AI, several limitations should be acknowledged. First, our sample was modest—17 digital illustrators—with inevitable attrition over time. Although such challenges are common in longitudinal research, they may affect representativeness. For instance, we observed that professional artists appeared to adopt AI more quickly than hobbyists, but this trend may not hold in a larger or more diverse population. Future studies should validate these patterns through broader, multi-tiered samples.

Second, the study was situated in the Chinese digital art context, which carries distinct structural and sociocultural dynamics. Features such as the industrialization of illustration, evolving copyright norms, and national policy shifts may have uniquely shaped participants’ anxieties and pragmatic turn toward AI. These factors constrain generalizability. Comparative work across cultural settings could illuminate how creators’ responses to generative AI vary under different institutional and market conditions. Third, our focus on digital visual artists leaves open how AI is received in other creative fields—such as music, literature, or fashion—where creators may prioritize different values. Musicians, for example, might emphasize originality or ownership, while designers may favor speed and efficiency. 

\revise{Moreover, the longitudinal design assumes conceptual stability across time. Yet as both AI technology and its surrounding discourse evolved rapidly, participants’ understanding of survey items may have shifted. Terms such as “AI-generated art” or “originality” likely carried different meanings in 2021 than in 2025. Future longitudinal research could employ adaptive instruments or reflexive interviews to track such semantic drift more systematically.}

\revise{A further limitation concerns retrospective and affective bias. Because participants reflected on their past emotions and practices during later interviews, their recollections and tone may have been influenced by current attitudes toward AI. While such reflections reveal important meaning-making processes, future work could complement them with real-time diary studies or longitudinal emotion tracking to capture affective dynamics with greater immediacy.}

Finally, although we traced shifts in attitudes and identity over time, we did not disentangle these changes from concurrent technological advances. As generative tools matured—through innovations like prompt engineering, style transfer, or controllable outputs—their influence likely evolved. However, we did not systematically examine how specific affordances shaped perception. Future work should integrate longitudinal interviews with targeted system evaluations to parse the respective contributions of tool design and sociocultural identity work.

These limitations, however, do not detract from the study’s contributions. Rather, they point to promising directions for future research: expanding sample diversity and cultural scope; examining AI’s role across creative domains; and probing the interplay of technical development and identity negotiation. Advancing along these paths will foster more holistic understandings of how generative AI reshapes creative practice—and guide the design of inclusive, adaptable, and sustainable human–AI systems in the future.

\section{CONCLUSION}
This paper presented a five-year longitudinal mixed-methods study of 17 digital painters in China, tracing how their attitudes toward generative AI shifted from resistance, to pragmatic adoption, and ultimately to reflective reconsideration. Our findings highlight not only these temporal trajectories but also the horizontal forces of peer influence, emotional dynamics, and persistent ethical concerns. By situating these insights within HCI, we contribute one of the first longitudinal accounts of creative practitioners’ evolving relationships with AI, advance a theoretical perspective of technology adoption as ongoing identity and value negotiation, and offer design implications for supporting control, diverse modes of collaboration, and the creative potential of failure. Together, these contributions deepen our understanding of how generative AI reshapes creative practice and point to future directions for designing systems that respect and sustain human creativity.

\section{Acknowledgments of the Use of AI}
In preparing this manuscript, we used ChatGPT in limited ways to support formatting and consistency. Specifically, the tool was employed to assist in generating Overleaf code for demographic tables and detailed tables in the Appendix, to help transform quantitative and qualitative interview materials into more readable Appendix formats, and to check grammar and terminology consistency across sections. The use of AI was restricted to these supportive tasks only. All authors take full responsibility for the content, analysis, and claims made in this paper. All data are real, and all perspectives and interpretations presented are those of the authors; no fabricated data or non-genuine content were produced through the use of large language models.

\section*{Acknowledgments}

We thank all participants for their time and commitment to this five-year longitudinal study.

\section*{Declaration of Interest}

The authors declare that they have no known competing financial interests or personal relationships that could have appeared to influence the work reported in this paper.

\section*{Author Contributions}

Yibo Meng led the conceptualization and design of the study and contributed to data curation, formal analysis, investigation, methodology, project administration, validation, visualization, and manuscript writing. Ruiqi Chen contributed to investigation, formal analysis, methodology, visualization, and manuscript writing. Zhuoran Lu contributed to supervision, validation, visualization, and manuscript writing. Shuai Ma contributed to supervision, validation, visualization, and manuscript review and editing. Chengxi Zang contributed to supervision, investigation, methodology, project administration, validation, visualization, and manuscript writing.All authors have read and approved the final manuscript and agree to be accountable for all aspects of the work.

\bibliographystyle{ACM-Reference-Format}
\bibliography{main}

\section{Appendix}

\section*{Appendix 1: Qualitative Interview Guide}

The following semi-structured interview guide was used to collect qualitative data. Questions were organized into four thematic blocks: (A) Background Information, (B) Experience with Generative AI, (C) Attitudes toward Generative AI, and (D) Forward-looking Reflections. Probes were used where appropriate to elicit more detailed responses.

\subsection*{A. Background Information}
\begin{enumerate}[label=A\arabic*]
  \item Age
  \item Years of artistic practice
  \item Living context: urban / rural
  \item Primary artistic domain (e.g., commercial/original/published illustration; concept design for characters/scenes/props; comics/graphic novels/animation; game art; digital art; graphic design; other)
  \item Professional status (e.g., freelance artist, employee, studio founder, part-time artist, student, unemployed)
  \item Client type (e.g., individual clients, galleries, publishers, small companies, medium/large companies, none)
\end{enumerate}

\subsection*{B. Experience with Generative AI}
\begin{enumerate}[label=B\arabic*]
  \item Familiarity with generative AI (e.g., no knowledge, basic awareness, entry-level user, intermediate user, advanced user)
  \item Have you ever used AI tools in your creative process?
  \item Which AI tools have you used?
  \item At which stages of creation have you used AI? (e.g., ideation, early references, sketching, composition, rendering, post-processing, direct image generation)
  \item How have you integrated AI content into your work? (e.g., inspiration only, redraw based on reference, modify base image, partial use, minor adjustments, direct adoption)
\end{enumerate}

\subsection*{C. Attitudes toward Generative AI}
\begin{enumerate}[label=C\arabic*]
  \item Do you think AI can significantly improve work efficiency?
  \item Has AI helped you overcome creative bottlenecks?
  \item Has AI lowered the technical barriers for artistic creation?
  \item Can someone proficient in AI be considered an “artist”?
  \item Do AI-generated works have artistic value?
  \item Is “AI art” a new category of art?
  \item Is emotional expression something that AI can never replace?
  \item Is it unethical to train models on data without authorization?
  \item Should artists have the right to decide whether their works are used for training?
  \item Should copyright of AI-generated works belong to the user?
  \item Should AI-assisted works be priced lower?
  \item Will generative AI have disruptive impacts on the industry?
  \item Is AI mainly a threat to lower-tier commercial art?
  \item Will proficiency with AI become an essential skill?
  \item Will the value of purely hand-drawn works increase?
  \item Do you feel excited and optimistic about AI’s development?
  \item Do you feel anxious or uneasy about the rise of AI?
  \item Do you feel pressured to learn new technologies because of AI?
\end{enumerate}

\subsection*{D. Forward-looking Reflections}
\begin{enumerate}[label=D\arabic*]
  \item What do you see as the greatest opportunities brought by AI?
  \item What are the greatest challenges AI poses (or is expected to pose) for artists?
  \item What do you see as the most urgent ethical and copyright issues, and what solutions would you propose?
  \item What expectations do you have for industry platforms?
  \item What would you like to say to artists who reject AI and to those who fully rely on it?
  \item Other open comments
\end{enumerate}

\section*{Appendix 2: Detailed Participant Information (2021--2025)}

\begin{table}[htbp]
\centering
\scriptsize
\setlength\tabcolsep{3pt}
\caption{Participant creative domains, client types, and professional status across five annual waves (2021–2025).}
\label{tab:participants_full}
\begin{tabular}{c|ccc|ccc|ccc|ccc|ccc}
\toprule
ID & \multicolumn{3}{c|}{2021} & \multicolumn{3}{c|}{2022} & \multicolumn{3}{c|}{2023} & \multicolumn{3}{c|}{2024} & \multicolumn{3}{c}{2025} \\
\cmidrule(lr){2-4} \cmidrule(lr){5-7} \cmidrule(lr){8-10} \cmidrule(lr){11-13} \cmidrule(lr){14-16}
 & Field & Client & Prof. & Field & Client & Prof. & Field & Client & Prof. & Field & Client & Prof. & Field & Client & Prof. \\
\midrule
1  & Game Art      & Large Co. & Y & Game Art      & Large Co. & Y & Game Art      & Large Co. & Y & Game Art      & Large Co. & Y & Game Art      & Large Co. & Y \\
2  & Illustration  & Small Co. & Y & Illustration  & Small Co. & Y & Illustration  & Small Co. & Y & Illustration  & Small Co. & Y & Concept Design & Private   & Y \\
3  & Concept Design& Private   & Y & Concept Design& Private   & Y & Concept Design& Private   & Y & --            & --        & --& --            & --        & -- \\
4  & --            & --        & N & --            & --        & N & Concept Design& --        & N & Concept Design& --        & N & Concept Design& --        & N \\
5  & Animation     & Large Co. & Y & Animation     & Large Co. & Y & --            & --        & --& Animation     & Large Co. & Y & Animation     & Large Co. & Y \\
6  & Game Art      & Large Co. & Y & Game Art      & Large Co. & Y & Game Art      & Large Co. & Y & Game Art      & Large Co. & Y & Game Art      & Large Co. & Y \\
7  & Illustration  & Small Co. & Y & Illustration  & Small Co. & Y & Illustration  & --        & N & Illustration  & --        & N & Illustration  & --        & N \\
8  & Game Art      & Large Co. & Y & Game Art      & Large Co. & Y & Game Art      & --        & N & Game Art      & --        & N & Game Art      & --        & N \\
9  & Animation     & Large Co. & Y & Animation     & Large Co. & Y & Animation     & --        & N & Illustration  & Small Co. & Y & Illustration  & Small Co. & Y \\
10 & Digital Art   & --        & N & Digital Art   & --        & N & Digital Art   & --        & Y & Game Art      & Large Co. & Y & Game Art      & Large Co. & Y \\
11 & Concept Design& Small Co. & Y & Concept Design& Small Co. & Y & Concept Design& Small Co. & Y & Concept Design& Small Co. & Y & Concept Design& Small Co. & Y \\
12 & Digital Art   & --        & N & Digital Art   & --        & N & Digital Art   & --        & N & Digital Art   & --        & N & Digital Art   & --        & N \\
13 & Digital Art   & --        & N & Digital Art   & --        & N & Digital Art   & --        & N & Digital Art   & --        & N & Digital Art   & --        & N \\
14 & Digital Art   & --        & N & Digital Art   & --        & N & Digital Art   & --        & N & Digital Art   & --        & N & Digital Art   & --        & N \\
15 & Graphic Design& Publisher & Y & Graphic Design& Publisher & Y & Graphic Design& Publisher & Y & Graphic Design& Publisher & Y & Graphic Design& Publisher & Y \\
16 & Animation     & Large Co. & Y & --            & --        & --& Animation     & Large Co. & Y & --            & --        & --& --            & --        & -- \\
17 & Digital Art   & --        & N & --            & --        & --& --            & --        & --& --            & --        & --& --            & --        & -- \\
\bottomrule
\end{tabular}
\end{table}

\end{document}